\documentclass[twocolumn,superscriptaddress,amsmath,amssymb]{revtex4}

\usepackage{graphicx}
\usepackage{dcolumn}
\usepackage{bm}
\usepackage{amsmath}
\begin{document}

\title{Revealing the ultrafast light-to-matter energy conversion before heat diffusion in a layered Dirac semimetal}

\author{Y.~Ishida}
\affiliation{ISSP, University of Tokyo, Kashiwa, Chiba 277-8581, Japan}

\author{H.~Masuda}
\affiliation{Department of Applied Physics, University of Tokyo, Hongo, Tokyo 113-8656, Japan}

\author{H.~Sakai}
\affiliation{Department of Applied Physics, University of Tokyo, Hongo, Tokyo 113-8656, Japan}
\affiliation{Department of Physics, Osaka University, Toyonaka, Osaka 560-0043, Japan}

\author{S.~Ishiwata}
\affiliation{Department of Applied Physics, University of Tokyo, Hongo, Tokyo 113-8656, Japan}
\affiliation{JST, PRESTO, Kawaguchi, Saitama 332-0012, Japan}
\affiliation{ISSP, University of Tokyo, Kashiwa, Chiba 277-8581, Japan}

\author{S.~Shin}
\affiliation{ISSP, University of Tokyo, Kashiwa, Chiba 277-8581, Japan}

\date{\today}

\begin{abstract}
There is still no general consensus on how one can describe the out-of-equilibrium phenomena in matter induced by an ultrashort light pulse. We investigate the pulse-induced dynamics in a layered Dirac semimetal SrMnBi$_2$ by pump-and-probe photoemission spectroscopy. At $\lesssim$1~ps, the electronic recovery slowed upon increasing the pump power. Such a bottleneck-type slowing is expected in a two-temperature model (TTM) scheme, although opposite trends have been observed to date in graphite and in cuprates. Subsequently, an unconventional power-law cooling took place at $\sim$100~ps, indicating that spatial heat diffusion is still ill defined at $\sim$100~ps. We identify that the successive dynamics before the emergence of heat diffusion is a canonical realization of a TTM scheme. Criteria for the applicability of the scheme is also provided. 
\end{abstract}

\maketitle
Dirac fermions in matter, as realized in graphene~\cite{Novoselov_Science04} and on surface of topological insulators (TIs)~\cite{Hasan_RMP10}, yield intriguing properties: Dirac fermions are highly mobile~\cite{Novoselov_Science04,QuOng_Science10} and can tunnel through infinitely high barriers~\cite{Novoselov_NPhys06,PKim_NPhys09}; in magnetic fields, they can gain non-trivial Berry's phase~\cite{Novoselov_Nature05,PKim_Nature05}; when contacted to a superconductor, Majorana modes may emerge~\cite{FuKane_PRL08,Wilczek_NPhys09}. Dirac materials are also usable as wide-band mode lockers that create ultrashort laser pulses of any color, owing to an ability to absorb light over a broad range of wavelengths~\cite{ZSun_ACSNano10,HZhang_APL12,Bonaccorso_NaturePhoton10}; Broad-band lasing may also be realized~\cite{Bonaccorso_NaturePhoton10,Li_PRL12,Gierz_NMat13,Zhu_SciRep15}. Investigations of the optical responses of Dirac fermions have opened pathways to take control of their charge~\cite{Neupane_PRL15,Sobota_PRL11,Ishida_SmB6}, spin~\cite{Mclver_NNano12,Jozwiak_NPhys13,Kastl_NCom15} and topological properties~\cite{Oka_PRB09,Gedik_Science13} by light. 

When founding the functions of optically non-equilibrated Dirac materials~\cite{ZSun_ACSNano10,HZhang_APL12,Bonaccorso_NaturePhoton10,Li_PRL12,Gierz_NMat13,Zhu_SciRep15,Neupane_PRL15,Sobota_PRL11,Ishida_SmB6,Mclver_NNano12,Jozwiak_NPhys13,Kastl_NCom15,Oka_PRB09,Gedik_Science13}, it becomes important to understand how the energy of light is transferred to matter and converted thereafter. In fact, there is still no counterpart for the well-established 2-, 3-, or 4-level rate equations that nicely describes the population inversions in conventional lasers. Applying the concepts in two-temperature model (TTM)~\cite{Allen_PRL87} and Rothwarf-Taylor model (RTM)~\cite{RT_Model} can be a viable approach. However, there is no general consensus to what extent, if at all, the models are suited for the description. Whether to approach from the metal side (TTM) or from the semiconductor side (RTM) is also a question, because Dirac materials are semimetals. Indeed, there are discussions that the TTM scheme is broken in graphite~\cite{Ishida_SciRep11} as well as in cuprates~\cite{Pashkin_PRL10}, while extensions of TTM also exist~\cite{ThreeTTM_PRB13,Perfetti_PRL07}. Meanwhile, continuous progress in pump-and-probe methods allows us to investigate the ultrafast phenomena in depth, thereby allowing us to test and develop our understandings from the experimental side. Angle-resolved photoemission spectroscopy (ARPES) implemented by the pump-and-probe method enables us to resolve the carrier dynamics in energy ($E$) and momentum ($k$) space, as done on TIs~\cite{Zhu_SciRep15,Neupane_PRL15,Sobota_PRL11}, graphene~\cite{Gierz_NMat13}, and cuprates~\cite{Perfetti_PRL07,Lanzara_TrARPES_NPhys11,Cortes_PRL11,Smallwood_PRB15,Ishida_Bi2212_SciRep15}, to mention a few.  

By means of time-resolved ARPES (TARPES), we study the ultrafastly-induced dynamics in SrMnBi$_2$, which is a new type of Dirac materials that has layers of quasi-two-dimensional Dirac fermions~\cite{JPark_PRL11,KWang_PRB12,GFChen_APL12,Boothroyd_PRB14,ValleyPol_PRL14}. Utilizing the high energy resolution and stability of our apparatus~\cite{Ishida_RSI14}, we investigated the carrier dynamics induced by various pump-power values ($P$'s). We also looked into the 100-ps time region, which has not been investigated in detail because the pump-induced variations become small. Around 1~ps, the electronic recovery slowed upon the increase of $P$. This is the behavior of the phonon-bottleneck effect expected in the TTM scheme, although opposite trends have been observed to date in graphite~\cite{Ishida_SciRep11} and in cuprates~\cite{Cortes_PRL11,Smallwood_PRB15}. In the 100-ps time region, an unconventional power-law cooling took place, indicating that the spatial diffusion of heat is still ill defined. Based on the results, we discuss that the chronology of the Dirac-fermion dynamics before the emergence of heat diffusion is a nice realization of the TTM scheme. Our study provides criteria for the applicability of the TTM scheme as well as its solid realization, and lays a basis for understanding the ultrafast dynamics in matter. 

\begin{figure}[htb]
\begin{center}
\includegraphics[width=8.6cm]{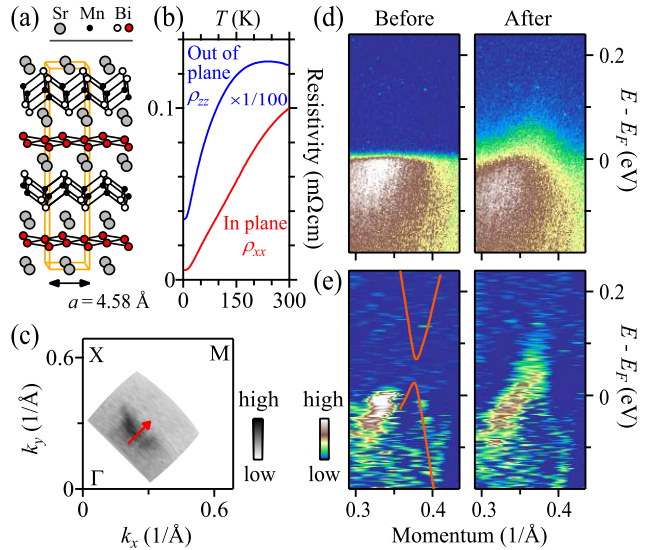}
\caption{\label{fig1}(Color online) Band dispersions of SrMnBi$_2$ extending into the unoccupied side. (a) Crystal structure of SrMnBi$_2$. The unit cell of 485~\AA$^3$ is indicated by a rectangular frame. (b) Temperature profiles of the resistivity. (c) Fermi-surface mapping. Spectral weight at $\pm$5~meV around $E_F$ is mapped in $k$ space. See supplementary movie file for the spectral-weight map from the occupied to unoccupied side~\cite{SOM}. (d) Band dispersions recorded before (-1.3~ps) and after (0.17~ps) the pumping ($P$ = 30~mW). The cut was along the Brillouin-zone diagonal indicated by an arrow in (c). (e) Second derivative of the images in (d) with respect to $k$. The calculated Dirac bands adopted from Ref.~\cite{GLee_PRB13} are overlaid on the left image. 
}
\end{center}
\end{figure}

SrMnBi$_2$ consists of alternate stacks of MnBi layers and Bi square nets separated by Sr ions [Fig.\ \ref{fig1}(a)]. The Bi square net has two Bi sites in its unit cell and hosts four anisotropic Dirac cones in the Brillouin zone~\cite{JPark_PRL11,XJZhou_SciRep14,HDing_PRB14}. The charge dynamics around $E_F$ is governed by the Bi electrons in the square net and is virtually decoupled from the Mn moments in the MnBi layers~\cite{JPark_PRL11,Boothroyd_PRB14}. Because the Dirac fermions exist in volume, the Dirac-related properties manifest as a bulk, such as the high bulk mobility and non-zero Berry's phase in the Shubnikov-de~Haas oscillations~\cite{JPark_PRL11}. Recent studies suggest that the Dirac fermions can be manipulated magnetically~\cite{BrianCS_PRB14,Masuda} and even modified into Weyl fermions~\cite{Borisenko} by replacing the Sr ions for other species.

SrMnBi$_2$ single crystals were grown by the self-flux method~\cite{Masuda}. The temperature profiles of the in-plane ($\rho_{xx}$) and interlayer ($\rho_{zz}$) resistivity [Fig.~\ref{fig1}(b)] displayed anisotropic metallic behavior characterized by $\rho_{xx}$(300~K)/$\rho_{xx}$(2~K) = 17.5 and $\rho_{zz}$(8~K)/$\rho_{xx}$(8~K) = 623. TARPES apparatus consisted of a hemispherical analyzer and a mode-locked Ti:Sapphire laser system delivering 1.48-eV pump and 5.92-eV probe pulses at 250-kHz repetition~\cite{Ishida_RSI14}. Time and energy resolutions were 270~fs and 18~meV, respectively. Samples were cleaved in the spectrometer at $\lesssim$5 $\times$ 10$^{-11}$~Torr. By utilizing a pin hole attached next to the sample, we estimated the spot diameters of the pump and probe beams to be 250 and 85~$\mu$m, respectively, and also checked that the movement of the pump beam was less than 5~$\mu$m when the delay stage was shifted for 600~ps pump-probe delay. Samples were held at $T_0 =$ 8~K during the measurements.

\begin{figure}[htb]
\begin{center}
\includegraphics[width=8.3cm]{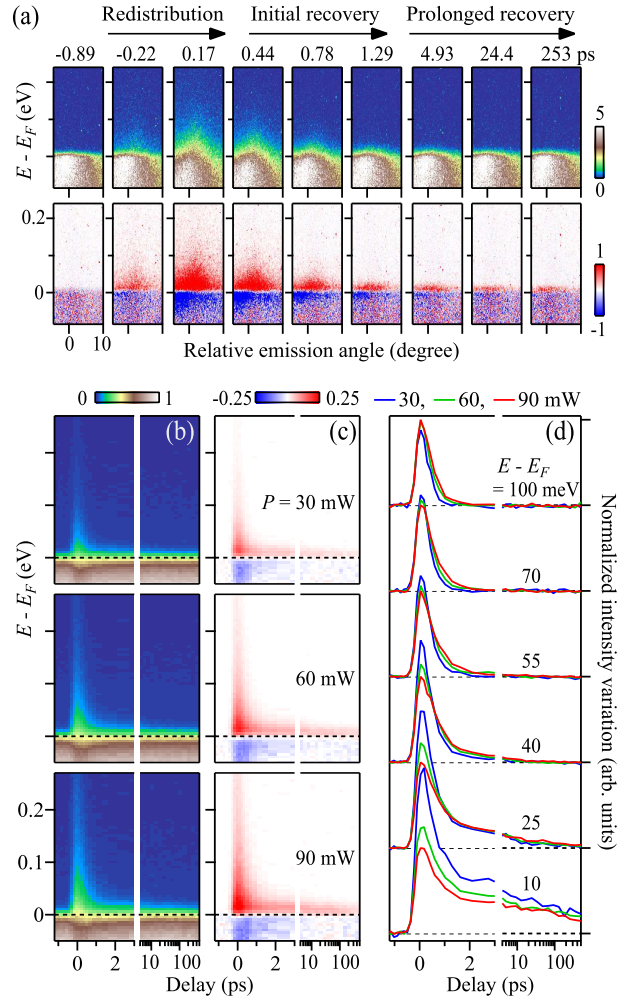}
\caption{\label{fig2}(Color online) TARPES of SrMnBi$_2$. (a) TARPES images. Top panels show band dispersions, and bottom panels show
difference to the averaged image before pumping. (b, c) Mappings of angle-integrated photoemission intensity (b) and its pump-induced difference (c) as functions of energy and delay time for three pump-power values $P$ = 30, 60, and 90~mW (top to bottom). $P$ = 30~mW corresponds to the pump fluence of 0.24~mJ/cm$^{2}$. (d) Photoemission intensity variation as functions of $t$ for the three pump-power values. }
\end{center}
\end{figure}

Figure \ref{fig1}(c) shows the results of the Fermi surface mapping. A crescent-like distribution is observed in the map as reported in previous ARPES studies~\cite{JPark_PRL11,XJZhou_SciRep14,HDing_PRB14}. Panels in Fig.\ \ref{fig1}(d) show band dispersions along $\varGamma$\,-\,$M$ [indicated by an arrow in Fig.\ \ref{fig1}(c)] recorded before (-0.80~ps) and upon (0.17~ps) the arrival of the pump pulse. In the latter, the unoccupied side is transiently populated. To highlight the bands~\cite{XJZhou_SciRep14,HDing_PRB14}, we present second derivative images with respect to $k$ in Fig.\ \ref{fig1}(e). We observe a conical band extending into the unoccupied side, which is attributed to the lower Dirac cone (LDC) crossing $E_F$; That is, the Dirac cone is $p$-type doped. The cone has a summit at 0.35~\AA$^{-1}$ away from $\varGamma$. The observed band nicely matched to the calculated LDC~\cite{GLee_PRB13} which has anisotropic band velocities of 2.1 and 2.8~eV\,\AA\, in the $\varGamma$ and $M$ sides, respectively, and that is gapped from the upper Dirac cone (UDC) due to spin-orbit interaction; see Fig.\ \ref{fig1}(e). Here, the calculated bands are overlaid on the image without any shift in $E$ and $k$. Pump-induced filling of UDC was negligibly small~\cite{SOM}, whose implication will be described later.

The electronic dynamics induced by the pump pulse was investigated by recording TARPES images along $\varGamma$\,-\,$M$ at various delays; see Fig.\ \ref{fig2}(a). In order to show clearly the pump-induced variations, we subtracted an average of 10 images recorded before the pumping ($t<$ -0.5~ps) from the TARPES images, and displayed the difference images in the lower panels. Upon the arrival of the pump pulse, the spectral intensity is spread into the unoccupied side. The changes are confined within $\sim$0.2~eV around $E_F$. Subsequent recovery is mostly accomplished within $\sim$2~ps, which is typical to the responses of metals~\cite{Fann_PRL92}. However, we also observe that the pump-induced changes remain even at  $>$100~ps. 

The dynamics is thus characterized by the following three stages: (1) A rapid pump-induced redistribution of electrons in a narrow energy range of $\sim$0.2~eV around $E_F$; (2) an initial recovery occurring within $\sim$2~ps; (3) a prolonged recovery of $>$100~ps. Below, we investigate the three stages step by step.

First, we look into the electron redistribution upon the arrival of the pump pulse. Panels in Figs.\ \ref{fig2}(b) and \ref{fig2}(c) respectively show angle-integrated photoemission intensity ($I$) and its pump-induced variation ($\varDelta I = I - I_0$) mapped in $\omega$\,-\,$t$ plane ($\omega \equiv E - E_F$). Here, $I_0(\omega)$ is the average of $I(\omega, t)$ at $t <$ -0.5~ps and represents the spectrum before pumping. At all pump power values investigated, we do not observe any delayed response, and the distribution nicely obeys the Fermi-Dirac function; See, supplementary movie file~\cite{SOM}. Shown in Fig.\ \ref{fig2}(d) are the intensity variation as functions of $t$ at various energy regions. Irrelevant to the pump power, we observe time-resolution-limited rise of intensity at $E > E_F$. 

The quasi-instantaneous realization of thermal electron distribution is typically observed in metals~\cite{Fann_PRL92}. The low-energy excitations across $E_F$ can screen the direct excitation, which, in effect, smears out the direct excitation and results in the electron redistribution around $E_F$ within our time resolution. In the case for TIs, there is a strong indirect-and-delayed filling of the surface Dirac bands, which is attributed to a process where electrons are first excited in the bulk across the band gap and then transferred into the surface bands~\cite{Sobota_PRL11}. Apparently, such a delayed channel does not exist in SrMnBi$_2$, and therefore, the metallic response dominated in the $p$-type-doped Dirac band. The absence of the filling into the UDC~\cite{SOM} further shows that the number of the electrons excited into the UDC was negligibly small. If such electrons had existed, they would have piled up at the bottom of the UDC to exhibit an inverted population, as observed in the surface Dirac bands of a $p$-type TI Sb$_2$Te$_3$~\cite{Zhu_SciRep15}. 

\begin{figure}[htb]
\begin{center}
\includegraphics[width=8.3cm]{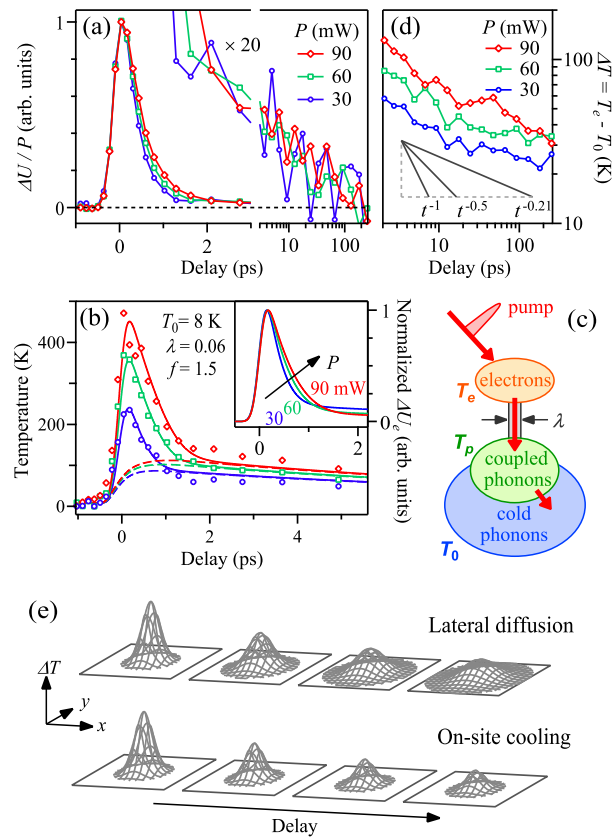}
\caption{\label{fig3}(Color online) Electronic recovery dynamics and its pump-power dependence. 
(a) Variation of the electronic energy $\varDelta U$ (see, text) normalized to the pump power $P$. (b) Variation of the electronic temperature $T_e$. The solid and dashed lines are numerical simulations of $T_e$ and $T_p$, respectively. Inset shows the calculated $\varDelta U_e(t)$ that are normalized to the peak heights. (c) Sketch of the energy flow in the TTM scheme. The conductance of the energy flow from the heated electrons to coupled phonons is described by the coupling constant $\lambda$. (d) Double logarithmic plots of $T_e$ versus $t$. Power-law exponents of -1, -1/2, and -0.21 are indicated by slopes. (e) Sketch of the lateral-diffusion cooling (top) and on-site cooling (bottom) after the sample is pumped by a finite pump-beam size. Note, the lateral spread of heat is absent in the latter. }
\end{center}
\end{figure}

Next, we investigate the initial electronic recovery within $\sim$2~ps. Here, we evaluate 
$\varDelta U(t) \equiv \int_{\omega\ge 0}\omega\varDelta I(\omega, t)\,d\omega$, which is a good measure of the excess electronic energy~\cite{Ishida_SciRep11}. Figure \ref{fig3}(a) displays $\varDelta U(t)$ normalized to the pump power $P$ for three representative $P$'s. The peak height of $\varDelta U(t)/P$ around $t$ = 0 is overlapped, indicating that the excess electronic energy deposited by the pump pulse is linear to $P$. Moreover, we find that the recovery slows upon the increase of $P$, as opposed to the case in graphite~\cite{Ishida_SciRep11}. The slowing is a piece of evidence for the phonon bottleneck effect taking place in the TTM scheme, which we describe below. 

The TTM scheme is often a starting point to understand the ultrafast dynamics in metals~\cite{Allen_PRL87,Perfetti_PRL07}. In this scheme, the pump pulse first raises the electronic temperature $T_e$ while leaving the lattice temperature $T_p$ low. Subsequently, the heated electrons transfer energy to the lattice. The conductance of the energy flow from the hot electrons to the coupled 
optical phonons is described by the coupling constant $\lambda$~\cite{Allen_PRL87,Perfetti_PRL07}. In applying the TTM scheme, we have derived $T_e(t)$ by fitting $I(\omega, t)$ to a Fermi-Dirac function multiplied by a linear density of states~\cite{SOM}. We set that 3$f$ phonon modes are coupled to the electronic system. The heat capacity of the coupled phonon modes $C_p(T_p) = f{\tilde{C_p}}$ ($\tilde{C_p}$ is the heat capacity for one atom per unit cell) is described by the Debye temperature $T_D$ = 310~K~\cite{Debye_APL12}. The use of Einstein model for $\tilde{C_p}$ did not change the discussion below~\cite{SOM}. We also treat the uncoupled phonons as a heat bath held at $T_0$ = 8~K, to which the coupled phonons transfer energy with a characteristic time $\tau$. The energy rate equation of the extended TTM~\cite{Perfetti_PRL07} reads, 
\begin{eqnarray*}
\frac{C_edT_e}{dt} & = & P\mathit{G} - \frac{3\lambda C_e\hbar \omega_0^2}{\pi k_B}\frac{T_e - T_p}{T_e},  \\
\frac{C_pdT_p}{dt} &= & + \frac{3\lambda C_e\hbar \omega_0^2}{\pi k_B}\frac{T_e - T_p}{T_e} - C_p\frac{T_p - T_0}{\tau}.
\end{eqnarray*}
Here, $k_B$ and $\hbar$ are Boltzmann and Planck constants, respectively; $C_e/T_e =$ 36.5~mJ/mol\,K$^2$~\cite{SpecificHeat_PRB11} is the electronic specific heat coefficient; $\mathit{G}(t)$ is a Gaussian function representing the temporal profile of the pump pulse; $\hbar\omega_0$ = $\sqrt[3]{\pi/6}\times k_B T_D$ = 21~meV is the Einstein optical-mode energy corresponding to the vibrations of Bi atoms~\cite{SOM}. 

The set of $T_e(t)$ for three values of $P$ is displayed in Fig.\ \ref{fig3}(b). The profiles of $T_e(t)$ are nicely reproduced by the theoretical curves of $(\lambda, f)$ = (0.06, 1.5). $f $ is smaller than the number of the atoms in the unit cell [8; see Fig.\ \ref{fig1}(a)], indicating that only a portion of the phonon modes are heated up by the hot electrons within $\sim$2~ps. The coupled modes are naturally attributed to some optical modes~\cite{Perfetti_PRL07}, while the rest are the cold phonons composed of acoustic modes and the uncoupled optical modes. In the inset to Fig.\ \ref{fig3}(b), we display the excess electronic energy $\varDelta U_e$ calculated by the model through the relationship $\varDelta U \propto \varDelta T_e^2$. The recovery of $\varDelta U_e$ slows upon the increase of $P$, and the model semi-quantitatively reproduces the pump-power dependence of $\varDelta U$ displayed in Fig.\ \ref{fig3}(a). The slowing occurs because the heat transfer is bottlenecked by the small coupling constant $\lambda$ = 0.06 $\ll$ 1. We note that a so-called ballistic diffusive term~\cite{Lisowski_APA04} was not needed to describe the dynamics at $\lesssim$2~ps.

We thus demonstrated the match between the experiment and theoretical model not only for a particular $P$ but for a set of $P$'s, and hence, the phonon-bottleneck effect manifesting in the pump-power dependency. The analysis showed that the Dirac fermions are weakly coupled into some optical modes. Whether the bottleneck effect appears in the pump-power dependency or not can be a good measure of the applicability of the TTM scheme.

Finally, we investigate the prolonged recovery after $\sim$2~ps ($\equiv t_0$) detected by our apparatus. Figure \ref{fig3}(d) shows double-logarithmic scale plots of the electronic temperature variations $\varDelta T = T_e - T_0$ versus $t$. For all the pump power values investigated, we observe that the cooling after $t = t_0$ obeys a power-law behavior $\varDelta T \propto t^{-\alpha}$ with $\alpha$ = 0.21\,$\pm$\,0.02. The power-law exponent is different from what is expected in a diffusive-type cooling: If lateral heat diffusion out of the pumped region was taking place [two-dimensional diffusion; see upper schematic of Fig.\ \ref{fig3}(e)], $\alpha = 1$, whereas in the case for one-dimensional diffusion of heat into bulk, $\alpha$ = 1/2~\cite{SOM}. If we force to fit the cooling behavior by taking into account the spatial spread of $\varDelta T$ at $t$ = $t_0$, the diffusion coefficient became unphysically large~\cite{SOM}. 
The results indicate that the spatial diffusion of heat is still ill defined at $\sim$100~ps. In the TTM scheme, the cooling can nevertheless occur without violating the energy conservation law, because the composite of the electrons and coupled phonons can transfer the excess energy to the cold phonons [Fig.\ \ref{fig3}(c)] still existing in the pumped region; That is, a local cooling can occur, see Fig.\ \ref{fig3}(e). The power-law exponent provides constraints to the microscopic mechanism of the unusual cooling, such as anharmonic decays of optical phonons into acoustic modes.

The hot spot created by the pump should eventually cool down by transferring heat to the surroundings, so that the unusual cooling at $\sim$100~ps should subsequently crossover into the usual ones accompanying spatial flow of heat. 
Generally, heat flow is carried by the acoustic modes, because their group velocities ($v_g \sim$ 1\,-\,5~km/s) are larger than those of the optical modes. The gradual population of the acoustic modes through the anharmonic decay at $\sim$100~ps will gradually trigger the heat to flow out of the hot spot. The flow into bulk will predominate the initial cooling, because the temperature gradient is steepest in the depth direction: Typical depth of the hot spot $L \sim$ 100~nm (equivalent to the optical penetration depth of the pump~\cite{EEMChia_APL14}) is smaller than the spot diameter of 250~$\mu$m. Once the acoustic modes are populated, subsequent drop of temperature will prevail with the time scale $L/v_g \sim$ 20\,-\,100~ps; Therefore, the $\sim$100-ps time region can be regarded as the transient where the heat is starting to flow, cross-overing into the $\gtrsim$1000-ps time region where cooling can indeed occur through various types of heat flows including the ballistic transmission of acoustic phonons~\cite{Siemens_NMater10,Minnich_PRL11,Hu_NNano15}.  

We demonstrated that the ultrafastly induced dynamics in SrMnBi$_2$ exhibits the bottleneck effect (slowing of the electronic recovery upon increasing $P$) at $\sim$1~ps and subsequent unconventional power-law cooling in the 100-ps time region. These characteristics were nicely described by a TTM scheme, in which the Dirac fermions are weakly coupled into some optical phonons. Our study provides a starting point to understand the ultrafast dynamics under stronger couplings and excitations: The strong couplings may result in the breakdown of the TTM scheme~\cite{Ishida_SciRep11,ThreeTTM_PRB13,Pashkin_PRL10,DalConte_Science12,14PRL_Graphite_Diffraction} or the band structures~\cite{Gierz_NMat13,Cortes_PRL11,Moos_PRL01} and concepts of quasi-particles may have to be seriously taken into account~\cite{Lanzara_TrARPES_NPhys11,Ishida_Bi2212_SciRep15}; A warm-dense matter will be reached at stronger excitations~\cite{04PRL_WarmDenseAu,09Science_WarmDenseAu,14PRL_WarmDense_Graphite}. A variety of pump-and-probe methodologies are needed to deepen insights into the strongly-correlated ultrafast phenomena. 

This work was supported by JSPS KAKENHI (Nos.\ 24224009, 25620040, 26800165 and 15K13332), MEXT (Photon and Quantum Basic Research Coordinated Development Program), Thermal \& Electric Energy Technology Inc.\ Foundation, Iketani Science and Technology Foundation, Nippon Sheet Glass Foundation for Materials Science, and JST PRESTO (Hyper-nano-space design toward Innovative Functionality).

\end{document}